\documentclass[manuscript]{aastex}

\usepackage{graphicx}
\usepackage{dcolumn}
\usepackage{paralist}
\usepackage{comment}
\usepackage{epsfig,graphicx}
\usepackage{multirow}
\usepackage{wrapfig}
\usepackage{float}

\slugcomment{To be sumitted to AJ.}

\shorttitle{Synthetic tracking using ZTF data}
\shortauthors{C.Zhai et al.}

\def\degsq{{\rm deg}^2}

\begin{document}

\title{Synthetic tracking using ZTF Long Dwell Datasets}

\author{Chengxing Zhai}
\affil{Jet Propulsion Laboratory, California Institute of Technology, 4800 Oak Grove Dr, Pasadena, CA 91109, USA}
\email{chengxing.zhai@jpl.nasa.gov}
\author{Quanzhi Ye}
\affil{IPAC, California Institute of Technology, 1200 E. California Boulevard, Pasadena, CA 91125, USA}
\author{Michael Shao, Russell Trahan, Navtej S. Saini, Janice Shen}
\affil{Jet Propulsion Laboratory, California Institute of Technology, 4800 Oak Grove Dr, Pasadena, CA 91109, USA}
\and
\author{Thomas A. Prince}
\affil{Division of Physics, Mathematics and Astronomy, California Institute of Technology, 1200 E. California Boulevard, Pasadena, CA 91125, USA}

\begin{abstract}
The Zwicky Transit Factory (ZTF) is a powerful time domain survey facility with a large field of view. We apply the synthetic tracking technique to integrate
a ZTF's long-dwell dataset, which consists of 133 nominal 30-second exposure frames spanning about 1.5 hours, to search for slowly moving asteroids down to approximately 23rd magnitude.
We found more than one thousand objects from searching 40 CCD-quadrant subfields, each of which covers a field size of $\sim$0.73 $\degsq$.
While most of the objects are main belt asteroids, there are asteroids belonging to families of Trojan, Hilda, Hungaria, Phocaea, and near-Earth-asteroids.
Such an approach is effective and productive. Here we report the data process and results.
\end{abstract}
\maketitle

\section{Introduction}
Aligning and Stacking up images to improve sensitivities has been performed some time ago to search for faint outer solar system objects\citep{Tyson1992,Cochran1995}. 
Relatively recently, we developed the synthetic tracking (ST) technique based on the same idea to search for near-Earth-objects (NEOs) by
1) acquiring frames fast enough to avoid streaks in a single frame 2) searching systematically for moving signals in post-processing\citep{Shao2014, Zhai2014}.
Sophisticated data preprocessing and massively parallel computation is the key to success.
The signals are not required to be detectable in individual frame and this is typically the case for detecting faint objects.
Massively parallel computation is needed to do a systematic search over a velocity grid covering a velocity range of interest.
We adopt GPU-aided computation to carry out this search by brute force. 

Synthetic tracking has applications at different time scales depending on the rate of motion of objects of interest.
The rule of thumb is to set exposure time so that the fastest motion does not streak, {\it i.e.}~the motion is less than the size of the point-spread-function (PSF) or one pixel depending on which one is larger.
For searching faint satellites and NEOs, modern sCMOS cameras with low read noise (typically 1-2e$^-$) are required to take frames fast enough
to freeze these objects in a single exposure, yet still limited by the sky background.

\begin{figure}[ht]
\epsscale{0.65}
\plotone{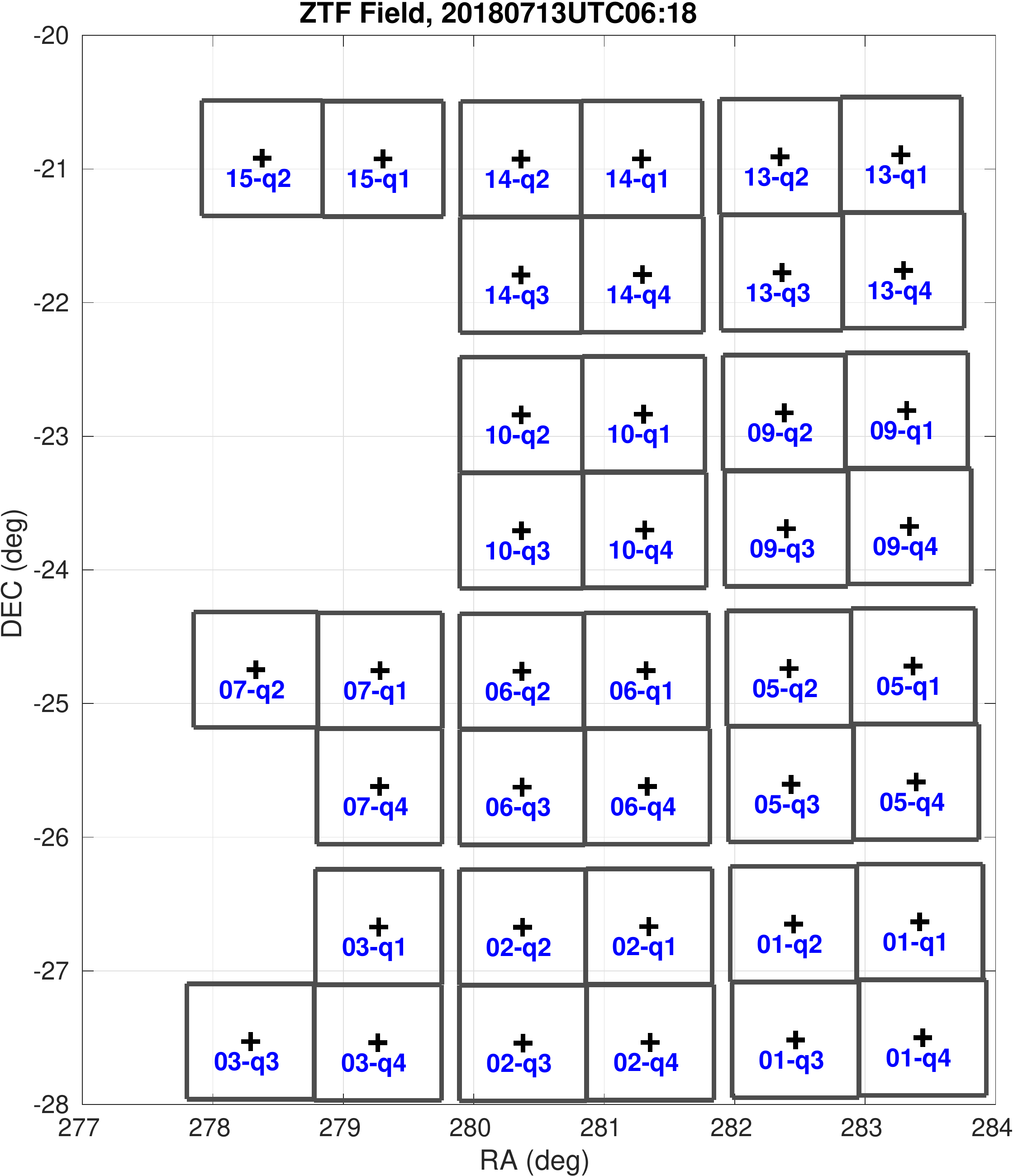}
\caption{ZTF field showing the coverage of the long dwell dataset taken on July 13, 2018. We only have data from 40 out of the total 64 CCD-quadrants. \label{ZTF_field}}
\end{figure}

The Zwicky Transit Factory is a new time-domain survey facility with 47 square degree field of view mounted on the Samuel Oschin 48-inch Schmidt telescope \citep{ZTF2018}. One of the goals of ZTF is to survey asteroids. The nominal data processing finds asteroids as significant signals in a single frame with exposure time of 30 seconds taken by CCDs \citep{ZTF2018,ZTF2019}.
Here we present results from applying synthetic tracking to a ZTF long dwell dataset, consisting of 133 30-second-exposure frames over about 1.5 hours, observing the same field.
Applying synthetic tracking to the long dwell data files from 40 CCD-quadrant sub-fields with each covering $\sim$0.73 $\degsq$ observing the galactic plane at [281, -24] deg in RA and DEC on July 13, 2018 (see Fig.~\ref{ZTF_field}), we found more than one thousand objects including main belt, Trojan, Hilda, Hungaria, Phocaea asteroids, and near-Earth-asteroids (NEAs). Stacking up 133 30-second images enables us to detect signals as deep as $\sim$23rd magnitude, while the detection threshold for each ZTF field is at $\sim$20.5 magnitude with the ZTF-r filter\citep{ZTF2018}. In the following sessions, we will describe the ZTF data and data process, present results, and then conclude with discussions on the future of this approach.

\section{Data and Method Description}
ZTF focal plane is covered by 16 CCDs, each having 4 independent amplifier channels, or CCD-quadrants, outputting digital signals of array size 3080${\times}$3072 pixels. The plate scale is $\sim$ 1 as/pixel,  so each quadrant covers a field of size  $\sim$0.73 $\degsq$. The nominal exposure time is set to 30 seconds.
The readout time is 8 seconds and the overhead is 2 seconds giving approximately a rate of 40 seconds per frame. The read noise is 10e$^-$ per read.
The detection limit per frame (30 second integration) is $\sim$ 20.5 mag for the ZTF-r band filter, limited by the sky background noise \citep{ZTF2018}.
The long dwell dataset that we analyzed was taken with the ZTF-r filter over the same field containing 133 frames,  giving total time span of $\sim$5200 seconds%
\footnote{Most of the CCD-quadrants have 133 frames, but a few CCD-quadrants have less frames. Among the 40 CCD-quadrants we processed, they all have more than 100 frames.}.
The data files belong to the ZTF data product called  {\it Epochal-difference image files} \citep{ZTF2018}, 
where for each CCD quadrant, a reference frame for the field was subtracted from each of the 30-second science frame. 
The science frame is obtained from the raw frame after applying bias, flat field, nonlinearity corrections. The streaks from aircraft and satellites and bright source halos and ghosts are masked out. The reference frame provides a static representation of the sky and is
generated from co-adding from 15 to 40 30-second images that pass specific data quality criteria.
The frames we used have all the static objects in the sky subtracted down to limiting magnitudes of the reference frames, which goes $\sim$1.5 mag deeper than individual 30-second images from averaging 15--40 images \citep{ZTF2018}. 

\begin{figure}[ht]
\epsscale{0.4}
\plotone{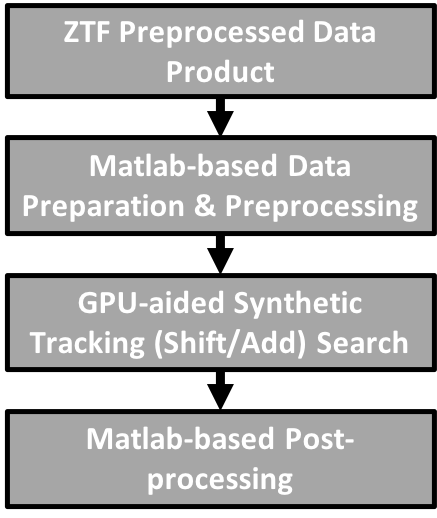}
\caption{Overall data processing flow chart.\label{flow}}
\end{figure}

\begin{figure}[ht]
\epsscale{0.7}
\plotone{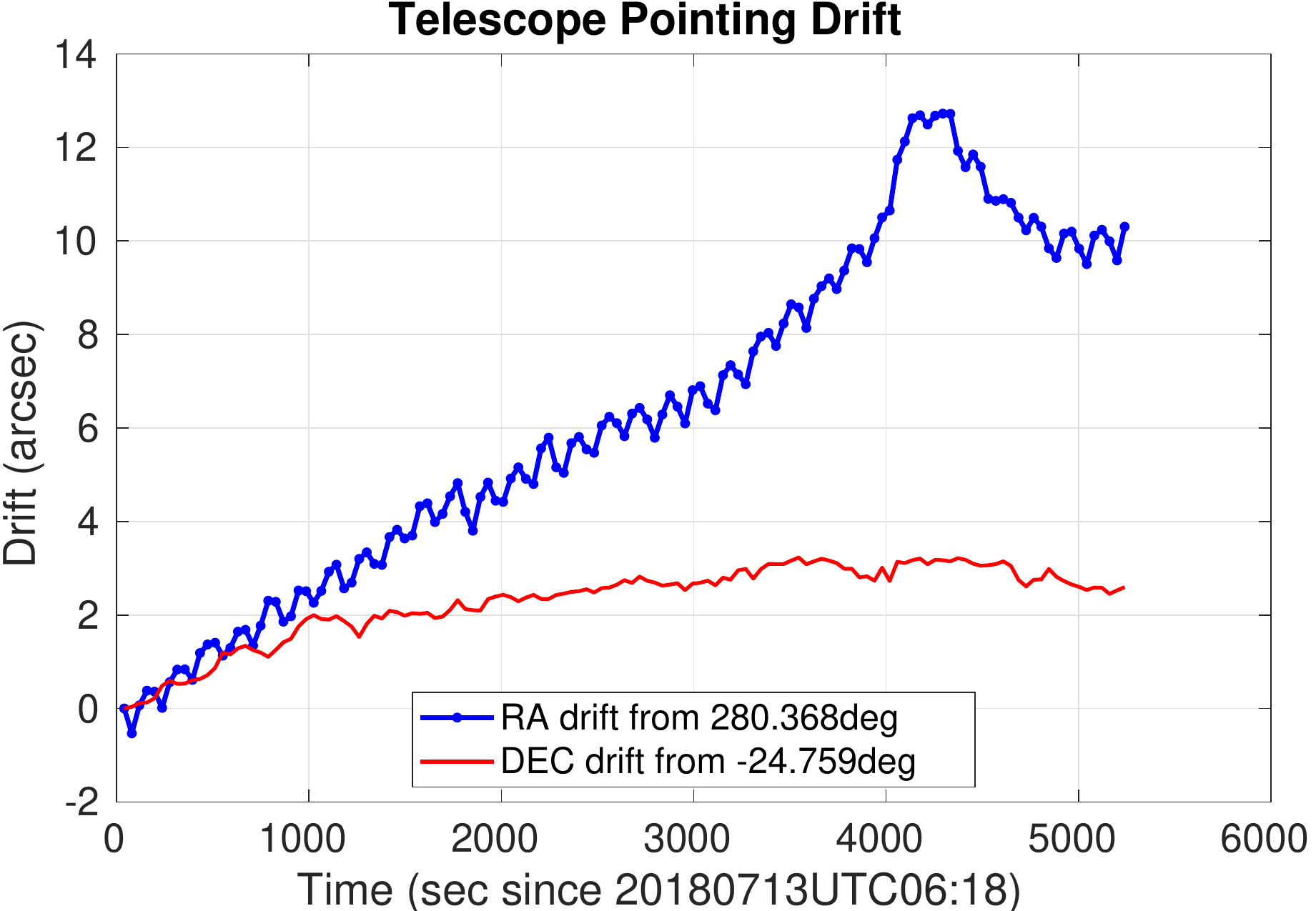}
\caption{ZTF pointing drift over $\sim$1.5 hours on July 13, 2018.\label{trackingDrift}}
\end{figure}

As summarized in Fig.~\ref{flow}, we start with the ZTF {\it Epochal-difference image files} and first use Matlab to prepare the data in the format suitable for our main synthetic tracking search engine to run. We re-register the frames by shifting an integer amount of pixels according to the sky coordinate of the center of each CCD-quadrant as recorded in the {\it fits header} data fields {\it CRVAL1, CRVAL2} to remove the drift of tracking (shown in Fig.~\ref{trackingDrift} )
so that all the 133 frames are aligned with respect to the sidereal. We then convert the ZTF data values into 16-bit unsigned integers, expected by our main synthetic tracking search engine. 
The ZTF preprocessing only subtracts out a static image of the sky for the field, so residual signals from static objects due to photon shot noise and atmospheric
effects are left in the frames. This is different from our standard synthetic tracking preprocessing, where we mask out star signals typically down to SNR 7
leaving much less static object residuals in the frames. 
Because of the higher level of residual noises, we do some extra preprocessing to set the pixels with high intensity values to the background value
to avoid generating too many false positives. There is quite some room left for improving the preprocessing to increase the detection sensitivity as discussed in section~\ref{sec:future}.

The main synthetic tracking search engine employs GPU-aided computation to perform shift/add in parallel integrating the frames
for all the trial velocities in a 100$\times$100 grid
with grid spacing of 0.25 pixel/frame covering a range of velocity of $\pm$0.75 deg/day in both RA and DEC.
A Gaussian kernel of FWHM = 3 pixel is applied to improve the signal-to-noise ratio (SNR)%
\footnote{Using a matched filter can improve this by about 5\%, but would require the details of the PSF shape.}.
For most of the CCD-quadrants, the detection threshold is SNR = 10, where the noise level is determined by the spatial noise (derived from spatial variations) in the frames, which varies between CCD-quadrants. However, a few CCD-quadrants yield too many false positives using detection threshold of SNR = 10,
so higher values like 15 or 20 are used instead. Signals above the detection threshold are found by the GPU-aided search and then clustered for post-processing.

In post-processing, each datacube (nominally 133 frames) is  broken into four sub-datacubes with equal number of frames.
Therefore, each sub-datacube would have nominally 33 frames. For a few CCD-quadrants that do not have all the133 frames (all have more than 100 frames),
each sub-datacube would have slightly less frames.
We require that the signal appears in at least 3 sub-datacubes at a significant level (above SNR=5), with PSF size within the range of $[1.2^{\prime\prime}, 3.5^{\prime\prime}]$, where
the PSF size is determined from fitting a Gaussian profile.
The field is quite crowded near the galactic plane with Galactic latitude -9 deg , so the asteroids have good chance to get close to a bright star's halo, which is masked out. A such example is shown in Fig.~\ref{partialObj}. Therefore, we only requires the signal appears in 3 sub-datacubes. 

For synthetic tracking search, we typically set a detection threshold much higher than the threshold used to detect signals in a single frame to avoid false positives from
integrating the frames in 100$\times$100 different tracking velocities \citep{Zhai2014}. With a detection threshold of SNR = 10, if the noises are close to Gaussian,
the false positive rate would be $\sim$ 3080$\times$3072$\times$100$\times$100$\times$erfc(10/$\sqrt{2}$)/2 $\sim$ 1e-12. However, the spatial noise is the dominant noise and most of them are the residual stellar signals, which is non-uniformly distributed. If the track of the signal goes through a region that has excessive stellar residuals, the detection signal could biased high.
This is the reason why we break the data cube into four sub-datacubes and required the signals to be in at least 3 sub-datacubes.
Because the object moves (the slowest object detected moves more than 6 arcsec per sub-datacube), it would be unlikely that all the sub-datacubes are biased high by excessive stellar noises.  Assume two sub-datacubes are not significantly biased by excessive stellar noise fluctuation, the false positive rate is 
$\sim$ 3080$\times$3072$\times$100$\times$100$\times$(erfc(5/$\sqrt{2}$)/2)$^2 \sim$ 0.0078. With the additional requirement of the appearance of the signal in the 3rd sub-datadube, we believe our false positive rate is negligible.

\section{Results}

\begin{figure}[ht]
\epsscale{1}
\plottwo{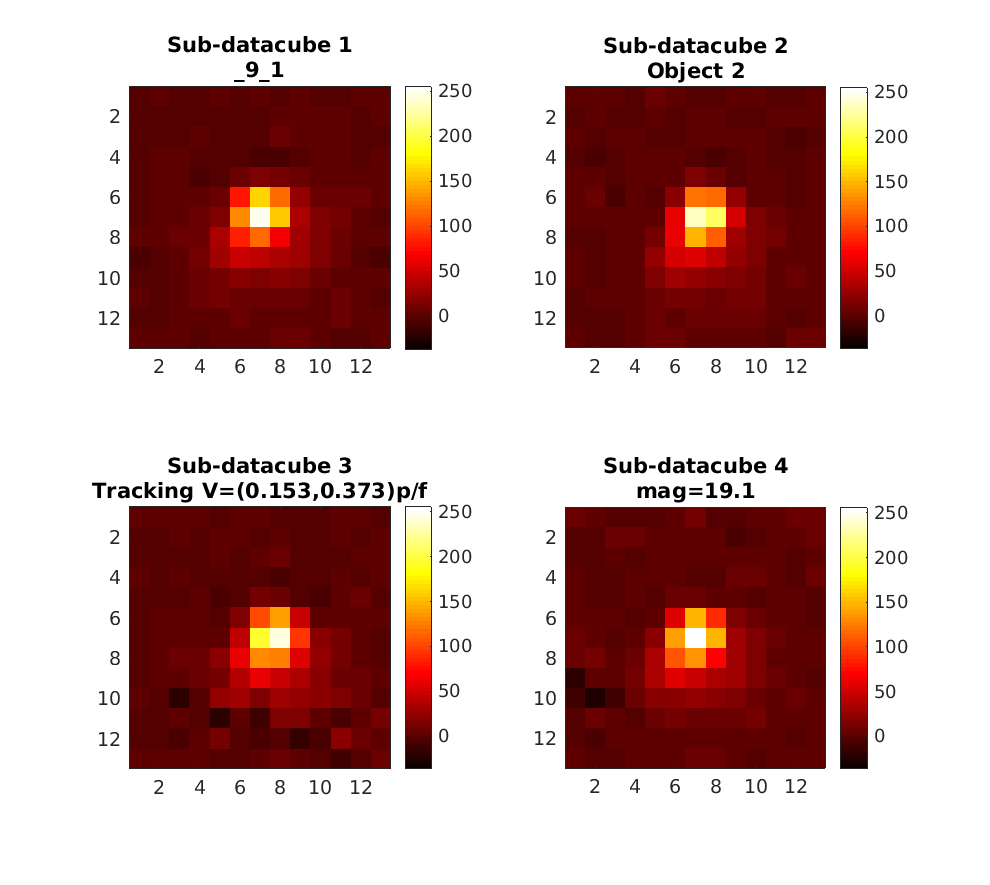}{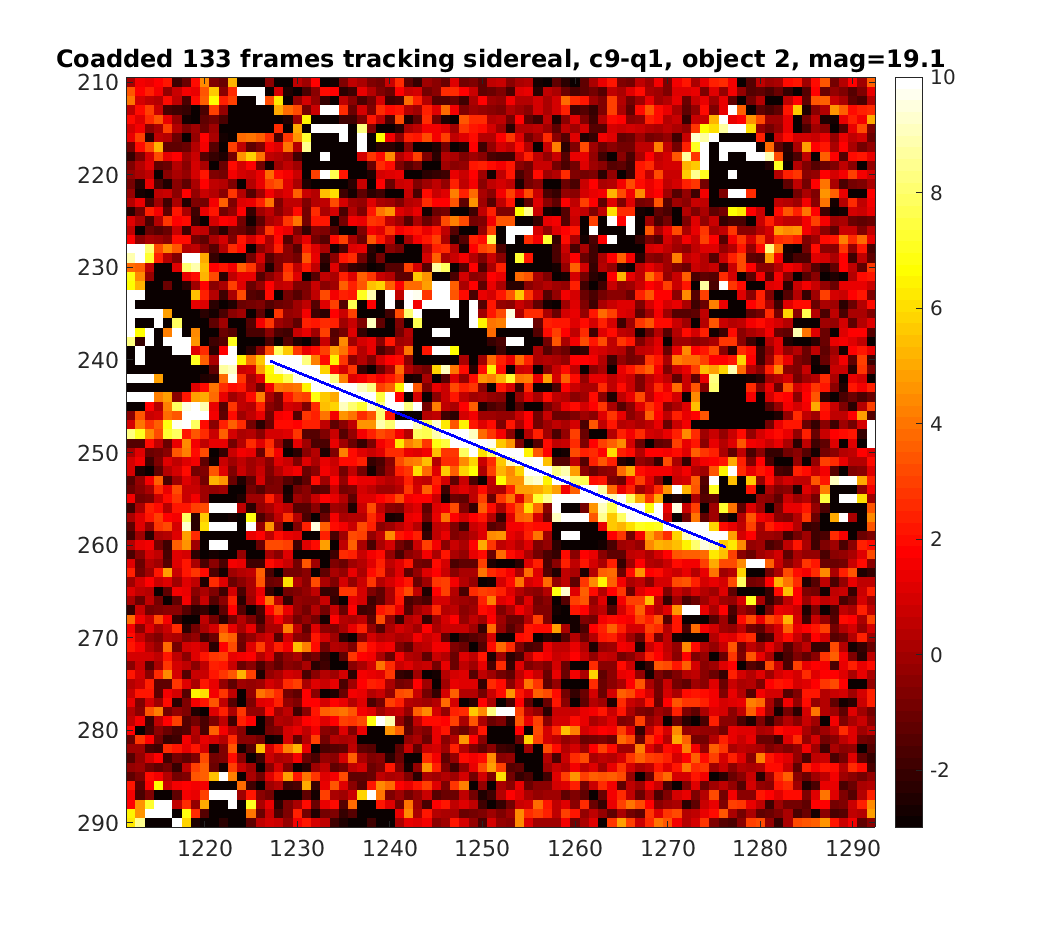}
\caption{Signal of a detected bright object in each subdatacube (left) and the streak in coadded image tracking sky (right).\label{brightObj}}
\end{figure}

\begin{figure}[ht]
\epsscale{1}
\plottwo{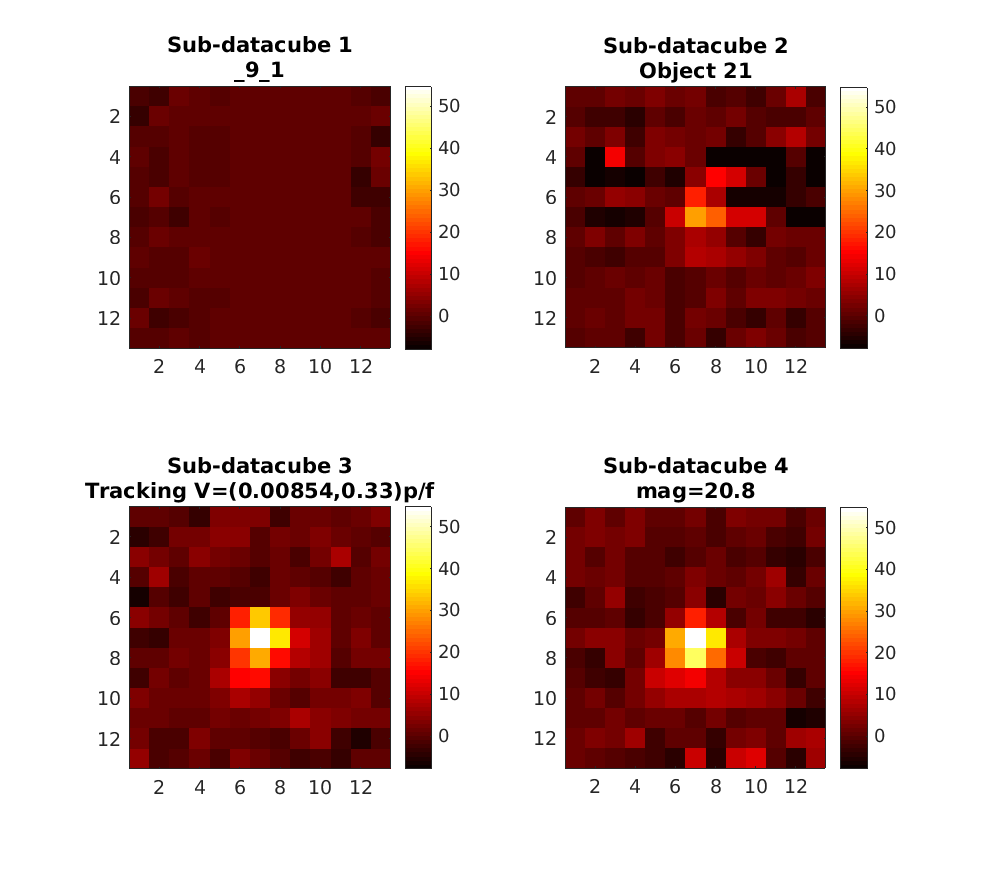}{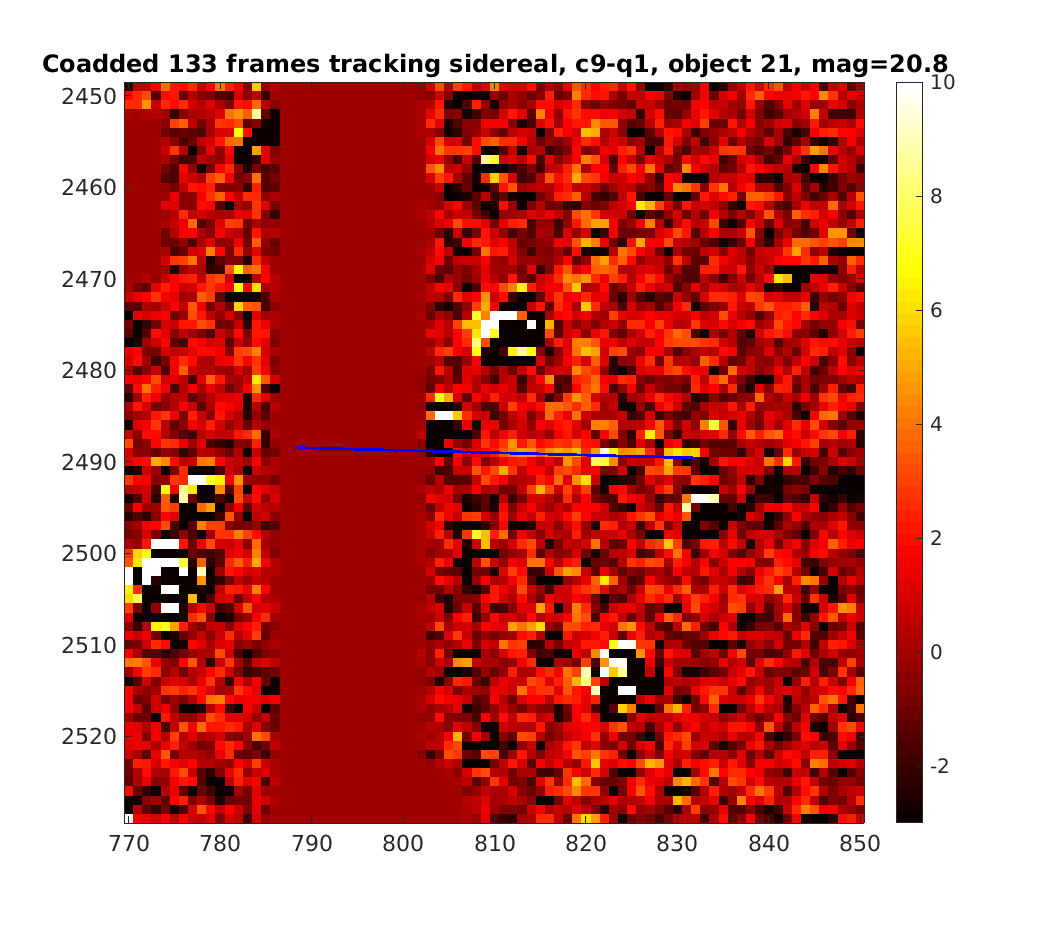}
\caption{Signal of a detected object with signal only in three sub-datacubes (left) and the streak running into a masked region in coadded image tracking sky.\label{partialObj}}
\end{figure}

\begin{figure}[ht]
\epsscale{1}
\plottwo{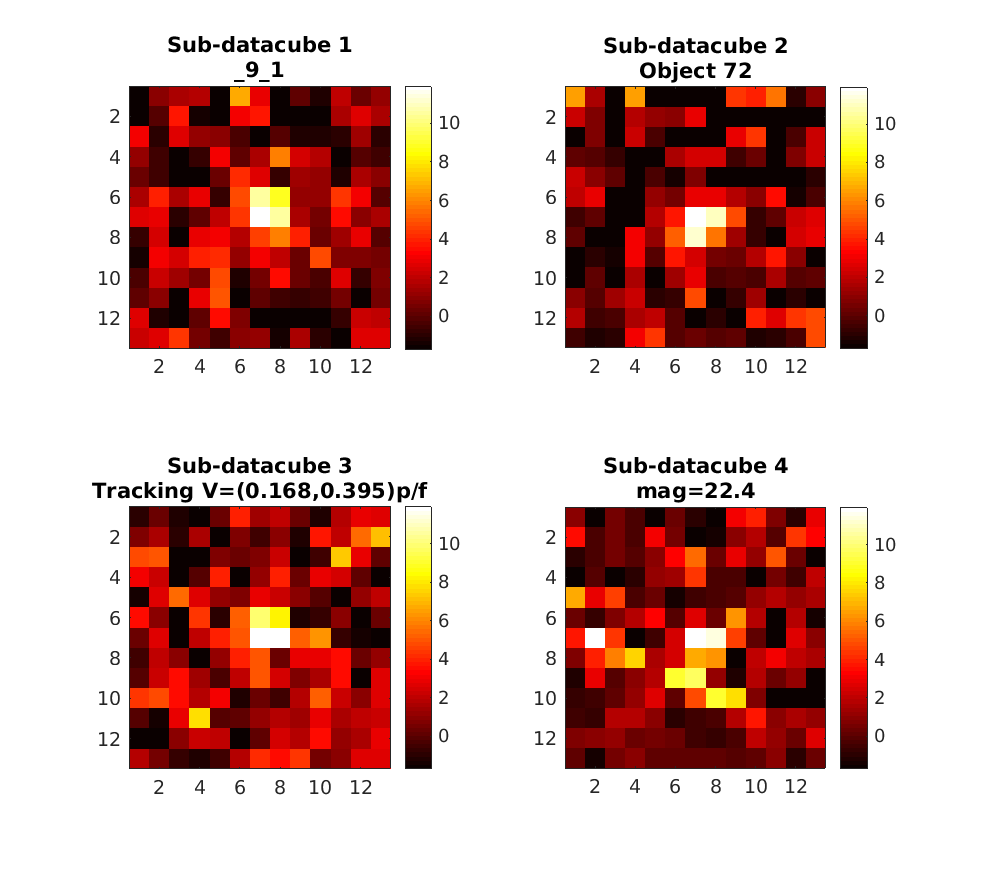}{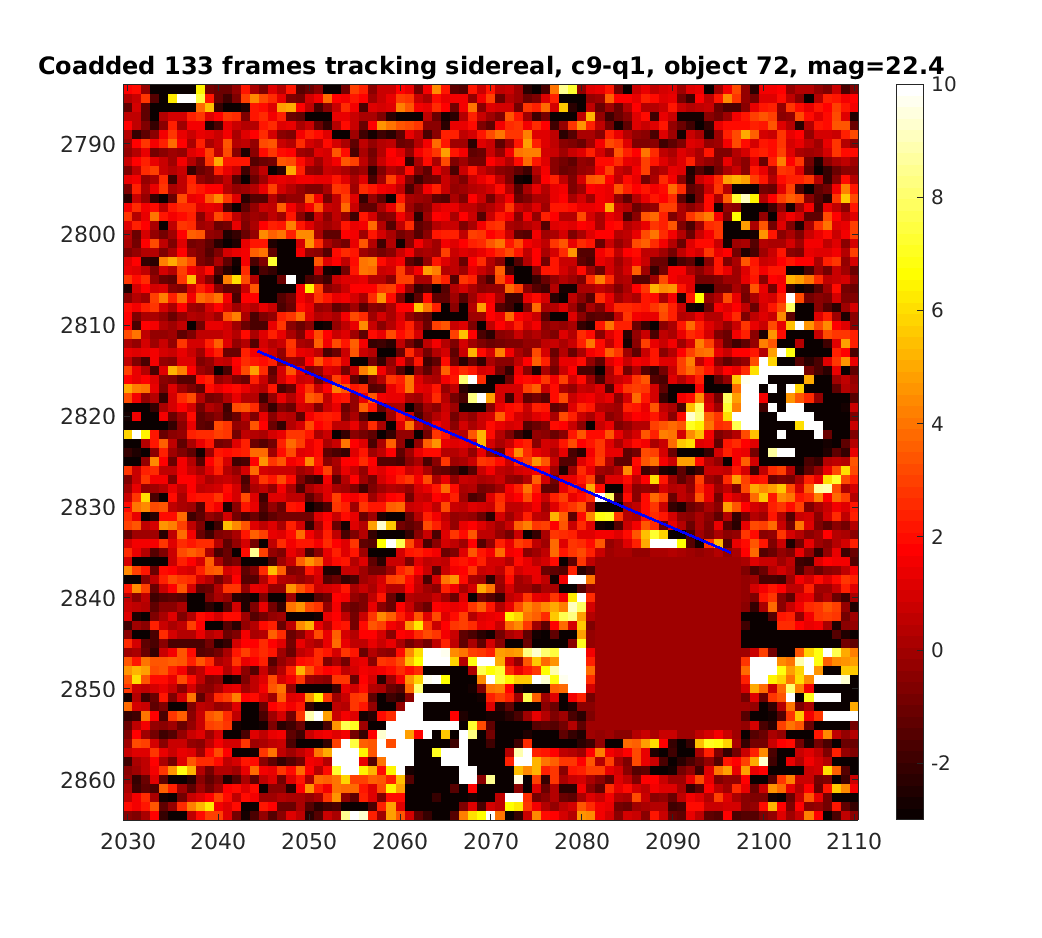}
\caption{Signal of a faint object with signal in all four subdatacubes (left) and the streak can not been seen in coadded image tracking sky.\label{faintObj}}
\end{figure}

In this section, we present examples of detected asteroid signals as well as statistics of the detections.
Left plot in Fig.~\ref{brightObj} displays the signals with color
scale showing in units of temporal noise level (estimated as temporal variation, {\it i.e.} frame-to-frame variation of pixel intensities) in four sub-datacubes when tracking the object at rate of [0.153, 0.373] pixel/frame, or $\sim$ [-35, -14] arcsec/hour, rate of a typical main belt asteroid, whose magnitude is 
estimated to be 19.1. It is bright enough to be detected as a streak as shown in the right plot in Fig.~\ref{brightObj} even with the trailing loss
when co-adding all the 133 frames to simulate an integration of $\sim$ 1.5 hours for ZTF tracking the sidereal.
Even though this can be detected as a streak in the co-added images, for accurate astrometry, the whole datacube is needed to take the advantage of
synthetic tracking \citep{Zhai2018}.
Fig.~\ref{partialObj} displays similarly another main belt asteroid of magnitude of 20.8, whose signals only show in three sub-datacubes
because it ran into a masked region due to a bright star. Note that the signal in the second sub-datacube is also slightly lower because
the signals of the asteroid in a few frames in the second sub-datacube are masked out.
Because the field is pretty crowded, it is not uncommon
for signals to be masked out, so we accept objects that has signals in three sub-datacubes. Even though the SNR of this object is $\sim$ 50 in a sub-datacube,
 the streak of this object is hard to identify and is no longer significant due to the trailing loss in the right plot of Fig.~\ref{partialObj} when tracking the sidereal.
Fig.~\ref{faintObj} displays in the same fashion an asteroid of magnitude of 22.4, whose signals consistently show in all four subdatacubes. Note that 
the color scale is measured in units of temporal noise level, the actual SNR with respect to spatial noise level is less than 10
because the signals are surrounded by spatial noises, which are higher than the temporal noise (unit of the color scale). The track of the object is marked in the right image of
tracking the sidereal, but signals are completely buried in the noises. This shows the efficacy of synthetic tracking.

\begin{figure}[ht]
\epsscale{0.8}
\plotone{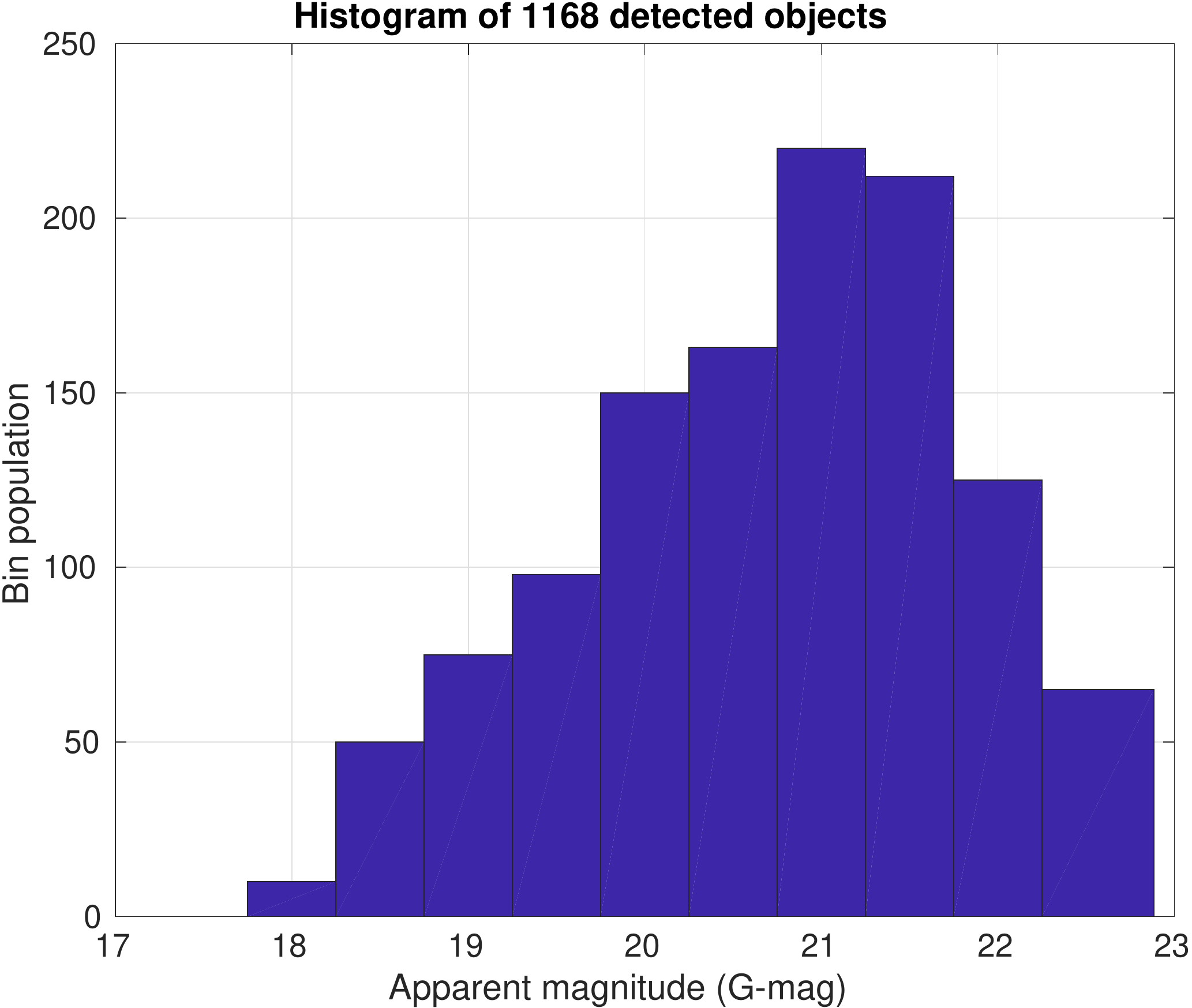}
\caption{Histogram of estimated magnitudes of detected objects.\label{mag_hist}}
\end{figure}

\begin{figure}[ht]
\epsscale{0.7}
\plotone{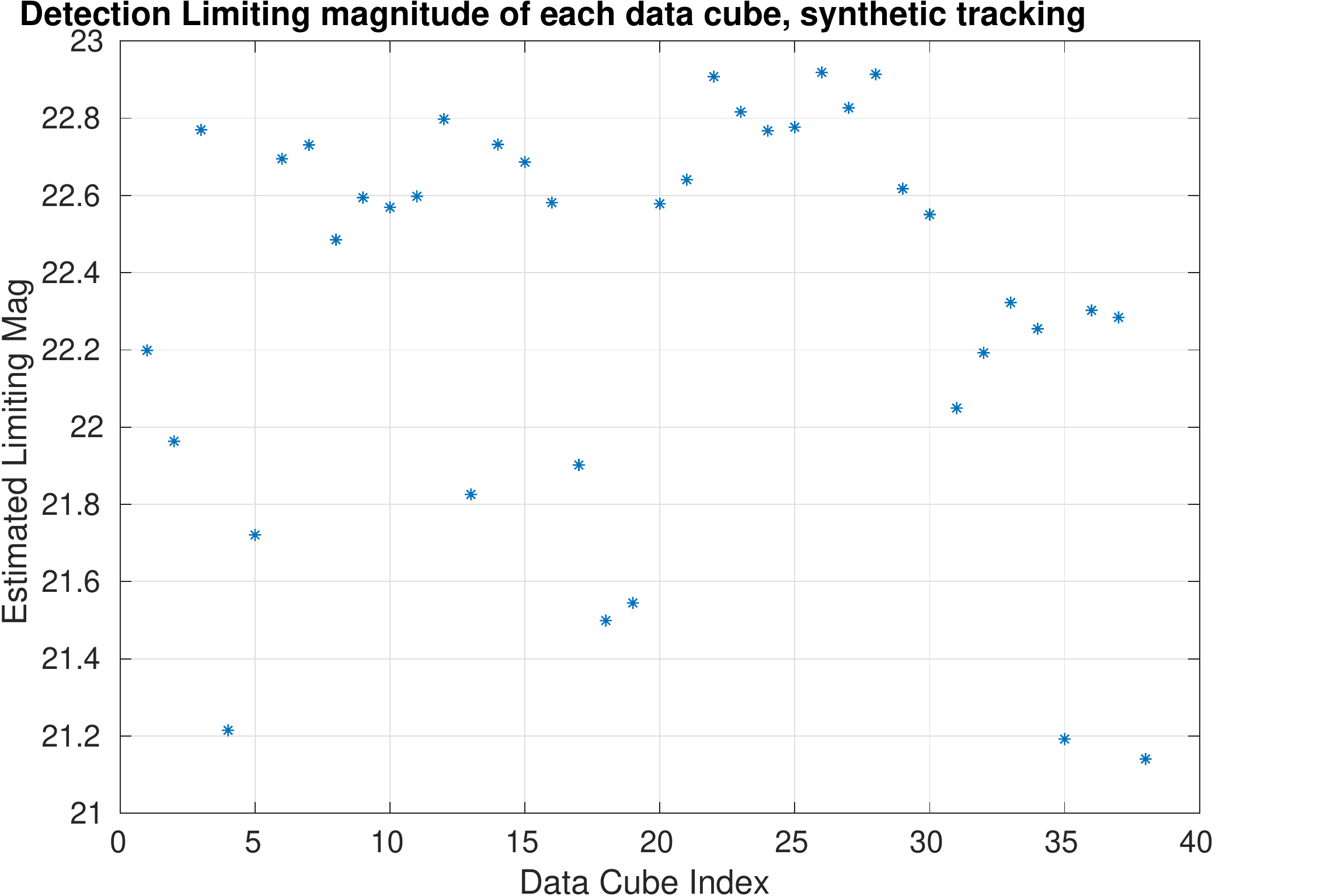}
\caption{Limiting magnitudes of data cubes.\label{limitMags}}
\end{figure}

Fig.~\ref{mag_hist} displays the histogram of the estimated magnitudes of the detected objects with bin size of 0.5 stellar magnitude.
We can see the increase of the population until beyond 21st magnitude, showing the increase of
the population of the asteroids with the apparent magnitude (as the sizes of asteroids get smaller). 
The drop of population in bins higher than 21st magnitude tells us the incompleteness of detection beyond 21st magnitude.
This is mainly due to the fact that the limiting magnitudes are different for different CCD quadrants as shown in Fig.~\ref{limitMags} for their different spatial noise levels, PSF sizes, and data processing settings to avoid excessive false positives. The lowest limiting magnitude is 21.2, consistent with
the fact that our detection is complete at magnitude of 21.
The incompleteness also comes from the criteria of requiring signals to be in at least three sub-datacubes because
fainter signals tend to be more likely to become not significant if the signals are partially masked out or weakened by the fluctuations of spatial noises.

\begin{figure}[ht]
\epsscale{0.9}
\plotone{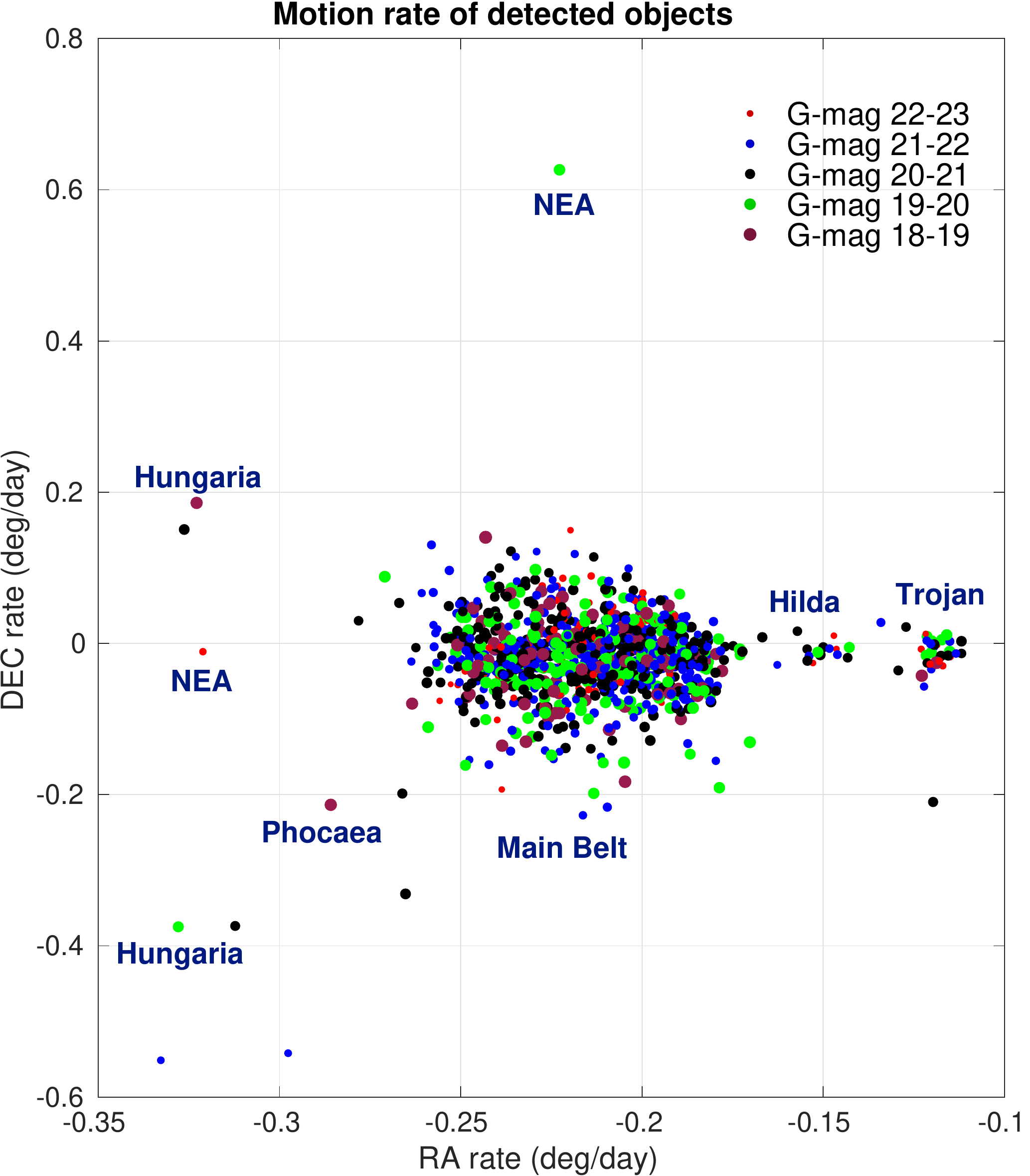}
\caption{Scatter plot of sky rate of the detected objects.\label{vel_dist}}
\end{figure}

The sky rate of the detected objects are displayed in Fig.~\ref{vel_dist}. Different colors and dot sizes label different ranges of apparent magnitudes
The field center was at [281, -24] deg, observed on July 13, 2019, UTC 6-7, so observations were approximately along the anti-sun direction.
Looking at the clusters in motion rate, we can easily identify asteroids of different families. 
The group near -[0.12, 0] deg/day are Trojan asteroids, with semi-major axis close to 5 AU.
The big cloud in the velocity distribution near ~ [-0.22, 0] deg/day are main belt asteroids. 
The small group between the main belt and Trojan asteroids is the Hilda asteroids with semi-major axis close to 4 AU. 
A few asteroids have RA rate in [-0.35, 0.3] deg/day with large DEC rates, which means their orbits having large inclinations.
They belong to the Hungaria family with semi-major axis between 1.78 AU to 2 AU and large inclination.
At least one asteroid belongs to Phocaea family between main belt asteroids and the Hungaria family with semi-major axis between 2.25 and 2.5 AU
and eccentricity greater than 0.1. In general, the sky rate of NEAs are not clustered and can have a wide range.
There are at least two known NEAs (2018 NE4, 2009 BC68), ``re-discovered'' with rates scattered outside the well-clustered groups. 
There should be quite some new asteroids detected here especially the ones fainter than 21st mag.
Using the MPCChecker (https://minorplanetcenter.net/cgi-bin/checkmp.cgi), we realized that fives asteroids 
( marked by the two blue dots with DEC rate $< $-0.5 deg/day, and three black dots with DEC rate $<$ -0.2 deg/day)
do not correspond to known asteroids. Some of them could be new NEAs.

\section{Conclusions and Discussions}
\label{sec:future}
We have presented the results from applying the synthetic tracking technique to a ZTF long dwell dataset to integrate over $\sim$1.5 hours.
We found more than one thousand asteroids including main belt, Trojan, Hilda, Hungaria, Phocaea family asteroids, and NEAs.
This approach is powerful and productive for surveying all types of asteroids.

The work presented here is a preliminary study. In the future, we can go deeper by masking out more residual noises from the static objects.
This can be done by using the raw ZTF data product and our standard data processing.
Furthermore, there are other ZTF fields that are less crowded, which would allow us to have less masked signals.
With better preprocessing, we expect to go deeper beyond 23rd magnitude and significantly increase completeness between magnitude range 21-23 for surveying
slowly moving objects. More than one thousand objects per 1.5 hour is a very good rate of yield and worth pursuing further.

\acknowledgments
We thank Frank Masci for information of ZTF data product.
Based on observations obtained with the Samuel Oschin 48-inch Telescope at the Palomar Observatory as part of the Zwicky Transient Facility project. ZTF is supported by the National Science Foundation under Grant No. AST-1440341 and a collaboration including Caltech, IPAC, the Weizmann Institute for Science, the Oskar Klein Center at Stockholm University, the University of Maryland, the University of Washington, Deutsches Elektronen-Synchrotron and Humboldt University, Los Alamos National Laboratories, the TANGO Consortium of Taiwan, the University of Wisconsin at Milwaukee, and Lawrence Berkeley National Laboratories. Operations are conducted by COO, IPAC, and UW.
This work has made use of data from the European Space Agency (ESA)
mission {\it Gaia} (\url{https://www.cosmos.esa.int/gaia}), processed by
the {\it Gaia} Data Processing and Analysis Consortium (DPAC,
\url{https://www.cosmos.esa.int/web/gaia/dpac/consortium}). Funding
for the DPAC has been provided by national institutions, in particular
the institutions participating in the {\it Gaia} Multilateral Agreement.
The work described here was carried out at the Jet Propulsion Laboratory, California Institute of Technology,
under a contract with the National Aeronautics and Space Administration.
Copyright 2019. Government sponsorship acknowledged.

\clearpage

\end{document}